\documentclass{PoS}

\usepackage[]{latexsym}
\usepackage[]{amsfonts}
\usepackage[]{amsmath}
\usepackage[]{graphicx}
\usepackage[]{epsf}

\newcommand{\ev}[1]{\langle #1 \rangle}
\newcommand{\Nc}{N_{\rm c}}

\def\lsi{\raise0.3ex\hbox{$<$\kern-0.75em\raise-1.1ex\hbox{$\sim$}}}
\def\gsi{\raise0.3ex\hbox{$>$\kern-0.75em\raise-1.1ex\hbox{$\sim$}}}
\newcommand{\lsim}{\mathop{\lsi}}
\newcommand{\gsim}{\mathop{\gsi}}

\title{Spectrum of SU(2) gauge theory with two fermions in the adjoint representation }

\ShortTitle{Spectrum of SU(2) gauge theory with two fermions in the adjoint representation }

\author{\speaker{Ari Hietanen}\\ Department of Physics, Florida International University, Miami, FL 33199, USA}

\author{Jarno Rantaharju, Kari Rummukainen\\
Department of Physics, University of Oulu, P.O.Box 3000, FIN-90014 Oulu, Finland} 

\author{Kimmo Tuominen\\
Department of Physics, University of Jyv\"askyl\"a, P.O.Box 35 FIN-40014 Jyv\"askyl\"a, Finland and Helsinki Institute of Physics, P.O.Box 64, FIN-00014 University of Helsinki, Finland}

\abstract{
 We present preliminary results of lattice simulations of SU(2) gauge
 theory with two Wilson fermions in the adjoint representation.
 This theory has recently attracted considerable attention because it might
 possess an infrared fixed point (or an almost-fixed-point), 
 and hence be a candidate for a walking technicolor theory.
 In this work we study the particle spectrum of the theory, and 
 compare it with more familiar spectrum of the theory with
 SU(2) gauge fields and two flavors of fundamental representation fermions.
}

\FullConference{The XXVI International Symposium on Lattice Field Theory \\
		 July 14 - 19, 2008\\
		 Williamsburg, Virginia, USA}

\begin{document}

\section{Introduction}

Gauge theories with fermions in other than fundamental representation
 may have qualitatively different features from QCD.  
One of these features is so called ``walking'' behaviour,
which is required by a class of technicolor theories \cite{TC}. 
Technicolor is an extension to 
the standard model, where the Higgs boson is replaced with 
a composite particle, essentially a scalar ``meson'' which consists
of techniquarks interacting through a gauge interaction, technicolor.  

In this project we study the properties of 
SU(2) gauge theory with two Dirac fermions in the adjoint (2-index
symmetric) representation.
This is a candidate theory for a ``minimal'' (i.e. simplest) 
walking technicolor \cite{Sannino:2004qp}.  For this to be viable,
the theory should either have an IR fixed point (where it
shows conformal behaviour) or an almost fixed point, where
the coupling evolves extremely slowly with the energy scale.
This theory has been studied
previously in 
\cite{DelDebbio:2008zf,Catterall:2007yx,Catterall:2008qk}.  
(See also \cite{Shamir:2008pb,Svetitsky:2008bw,DeGrand:2008dh,Fodor:2008hn} 
for recent studies of related theories.)

In this first stage we shall investigate the excitation spectrum of the theory
and estimate its lattice phase diagram. We use considerably
larger volumes than the earlier published work.  Full results
will be published in \cite{MWLattice}.

\section{Lattice action}

The lattice action of SU(2)+adjoint quark theory is
$
  S = S_{\rm g}+S_{\rm f},
$
where $S_{\rm g}$ is the standard plaquette gauge action and 
$S_{\rm f}$ is the Wilson fermion action for spinors in the
adjoint representation:
\begin{align}
  S_{\rm f} &= a^4\sum_x\bar{\psi}(x)D \psi(x) \\
  & = \sum_x\bar{\psi} \psi(x)-\kappa\sum_{\mu}\left[\bar{\psi}(r-\gamma_\mu)V_\mu(x)\psi(x)+\bar{\psi}(r+\gamma_\mu)V_{\mu}^\dagger(x-\mu)\psi(x-\mu)\right].
\end{align}
Here the adjoint link variables $V$ are related to the fundamental 
representation ones as 
\begin{equation}
  V_\mu^{ab}(x)=2{\rm{Tr}}(S^aU_\mu(x)S^bU_\mu^\dagger(x)),
\end{equation}
where $S^a = \frac12 \sigma_a$, $a=1,2,3$ are the generators of the 
fundamental representation.  
%Note that $V^{ab} \in R$.
%As we are going to study the SU(2) we assume from now on $a=1$, 2, 3. 

As usual, the lattice action is parametrized with
\begin{equation}
  \beta = \frac{2\Nc}{g^2} = \frac{4}{g^2}
\mbox{~~~and~~~}
%\end{equation}
%and
%\begin{equation}
  \kappa = \frac{1}{8+2m_{Q,\rm{bare}}},
\end{equation}
where $m_{Q,\rm{bare}}$ is the bare mass parameter.

\section{Lattice simulations}

The simulations were carried out with five different values of $\beta=1.3$, 1.7, 1.9, 2.2 and 2.5. For each value of $\beta$ we used 5 to 11 different values of $\kappa$. The volumes used were $24^4$ and $32^4$. The updates were performed using standard hybrid Monte Carlo algorithm and the number of trajectories was 100-700 for a single run. The timestep $\Delta \tau$ used was $0.02$ for larger values of mass and was decreased to $0.004$ closer to the zero mass limit. The number of integration steps $N_s$ was chosen so that $N_s \times \Delta\tau \sim \mathcal{O}(1)$.

We also performed some simulations in fundamental representation in order to 
validate the algorithms and to confirm that we observe qualitatively 
different behaviour and not e.g. lattice artifacts.

\subsection{Phase diagram}

First, in order to obtain the relevant parameter range we probe the phase
 diagram of the theory. We are especially interested in the critical line 
$\kappa_c(\beta)$, along which the quark mass vanishes.
%in $\beta$,$\kappa$ plane, where the physical quark mass is zero, which 
The quark mass is measured using the axial Ward identity (``PCAC mass''):
\begin{equation}
  m_{\rm Q}=\lim_{t \rightarrow \infty}\frac{1}{2}\frac{\partial_t V_{\rm PS}}{V_{\rm PP}},
\end{equation}
where the currents are
\begin{align}
  V_{\rm PS}(x_0) &= a^3\sum_{x_1,x_2,x_3} \ev{\bar{u}(x)\gamma_5d(x)\bar{u}(0)\gamma_5d(0)} \\
  V_{\rm PP}(x_0) &= a^3\sum_{x_1,x_2,x_3} \ev{\bar{u}(x)\gamma_0d\gamma_5(x)\bar{u}(0)\gamma_0\gamma_5d(0)}.
\end{align}

\begin{figure}
  \begin{center}
    \includegraphics[width=0.45\textwidth]{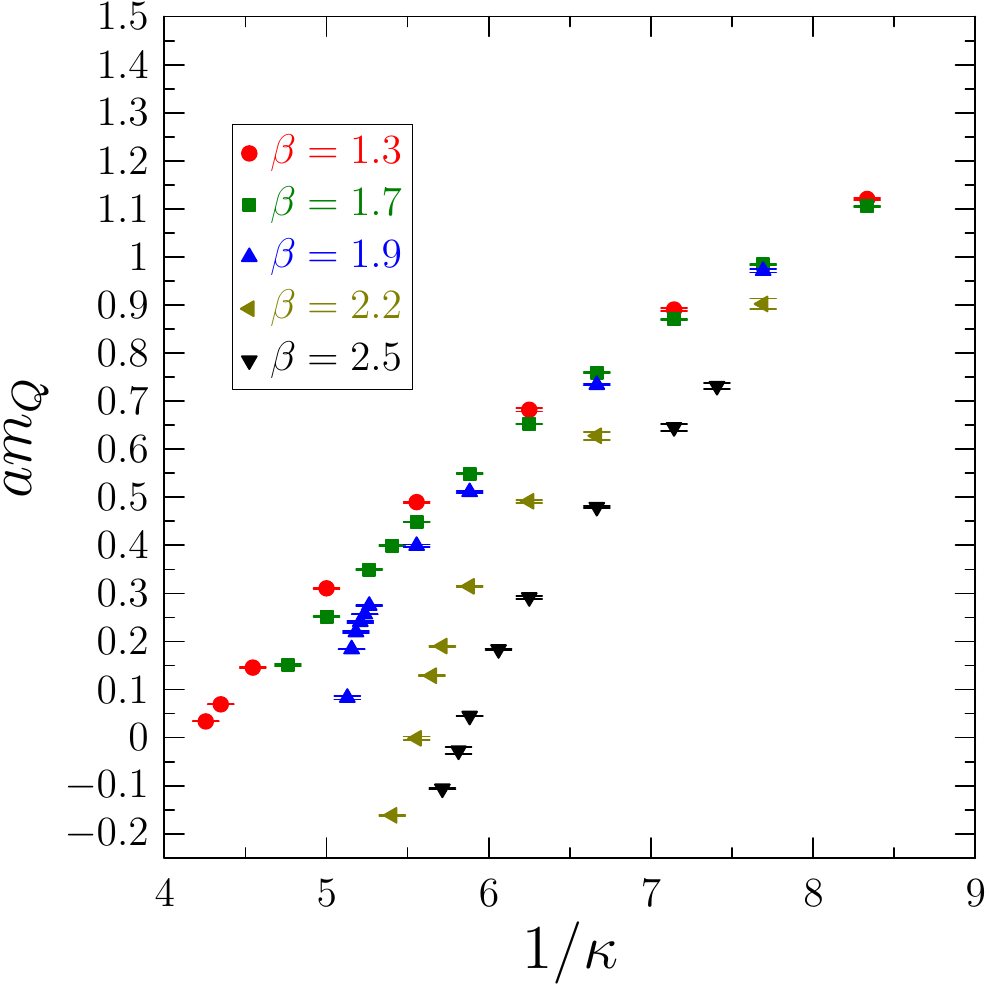}
    \includegraphics[width=0.45\textwidth]{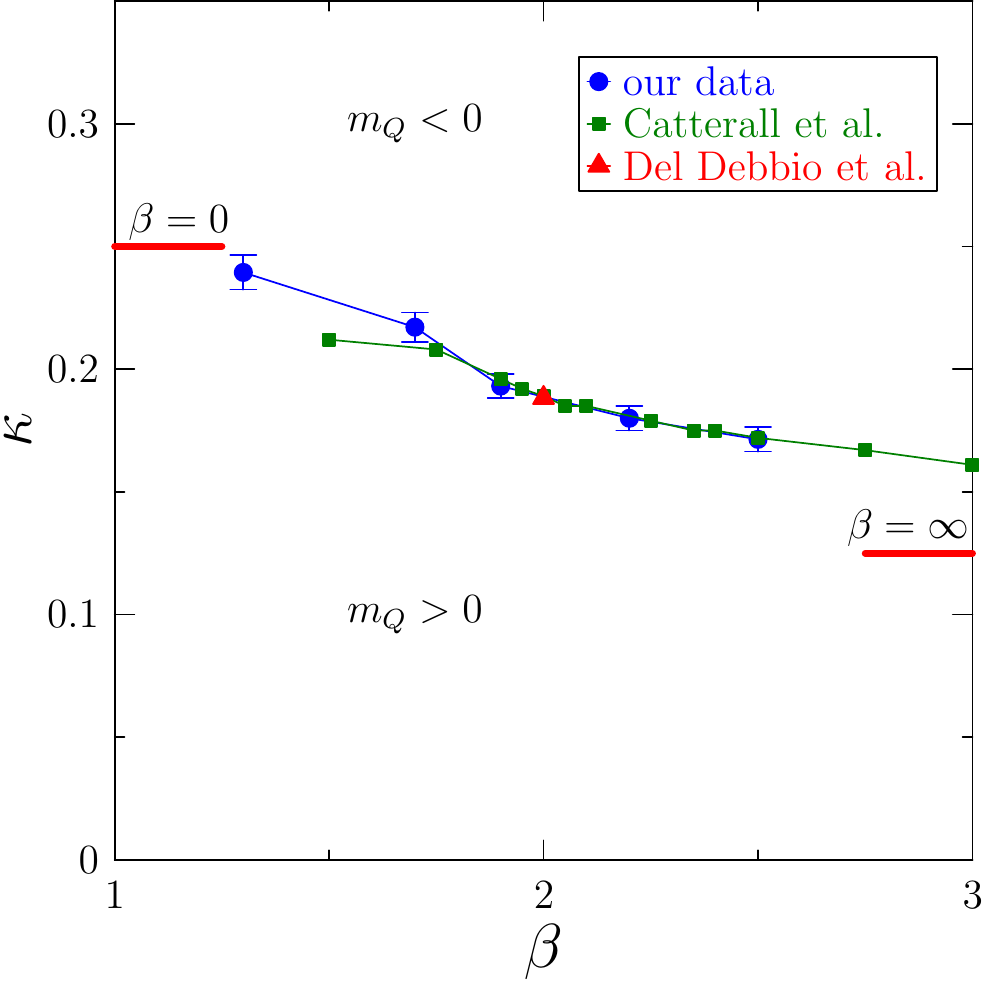}
    \caption{Left: Quark masses as functions of $\kappa$. 
  Right: The phase diagram with $m_{\rm Q}=0$ line. 
  Our results agree well with other groups.
      \label{phasedia}}
  \end{center}
 \end{figure}

On the left panel of Fig.~\ref{phasedia} we have plotted the measured quark 
masses against $1/\kappa$, and
on the right panel we have extrapolated the results to the zero quark mass 
limit. We include here also the results from Del Debbio 
et al.~\cite{DelDebbio:2008zf} and Catterall et al.~\cite{Catterall:2008qk}. 
All of the results agree well with each other.

Authors of \cite{Catterall:2008qk} 
find a phase transition around $\beta=1.9$ at zero mass.  We also observe
a change in the behaviour of the system at $\beta\approx 1.9$: 
when $\beta \lsim 1.9$ and we decrease the quark mass ($\kappa$ is increased),
 the system has a sharp phase transition around $m_{\rm Q} \approx 0$,
making the simulations in practice impossible at large volumes.  
This is especially noticeable at $\beta = 1.9$ simulations.  
This new phase is an artifact of Wilson fermions, corresponding to
the ``Aoki phase'' of lattice QCD.  
On the other hand, when $\beta \gsim 1.9$ there is no sharp transition and, if
the volume is not too large, we can decrease $m_{\rm Q}$ below zero without
problems.  

\subsection{Mass spectrum}

\begin{table}
  \begin{center}
    \begin{tabular}{|c|c|c|}
      \hline
      Particle & quark content & QCD equivalent\\
      \hline
      Pseudoscalar meson & $U\gamma_5\bar{U}$ & $\pi$\\
      \hline
      Vector meson & $U\gamma_\mu\bar{U}$ &  $\rho$  \\
      \hline
      Axial vector meson & $U\gamma_\mu\gamma_5U $ & $b_1$\\
      \hline
      ``Higgs'' & $U\bar{U}+D\bar{D}$  & $f_0$ or $\sigma$ \\
      \hline
      Spin 1/2 baryon & $UUD$ & proton \\
      \hline
      Spin 3/2 baryon & $UUU$ & $\Delta$ \\
      \hline
      Quark Gluon & $UG$ & NA\\
      \hline
    \end{tabular}
    \caption{List of some particles in adjoint representation \label{particles}}
  \end{center}
\end{table}

The ``hadron'' spectrum with adjoint representation quarks 
has more states than with the fundamental representation quarks.    
In Table \ref{particles} we have listed some of the low energy states.  Here we will present the measurements of the pseudoscalar and vector 2-quark
states, ``mesons'', 
and the spin 1/2 and spin 3/2 3-quark states, ``baryons''.

The measurement of the axial vector would be interesting, 
because it has been speculated that ratio 
$M_{Axial}/M_{\rho}$ could be smaller than one for a conformal 
theory~\cite{Appelquist:1998xf}.  Unfortunately, we do not have yet 
good enough statistics at large volumes to obtain reliable measurement. 
The ``Higgs'' particle includes a quark disconnected part and we have not 
attempted to measure it.

The masses of the excitations are estimated by fits to the time 
sliced averaged correlation functions.  We use wall sources at timeslice 
$t=0$ (with Coulomb gauge fixing) and point sinks.  For example, 
the correlation function for mesons reads
\begin{equation}
G_{\mathcal{O}}(t) \propto 
\sum_{x,y_1,y_2}\langle\bar{\psi}(x,t)\Gamma_{\mathcal{O}}\psi(x,t)
\bar{\psi}(y_1,0)\Gamma_{\mathcal{O}}\psi(y_2,0)\rangle,
\end{equation}
where $\Gamma_{\mathcal{O}}=\gamma_5$ for the pseudoscalar and $\Gamma_{\mathcal{O}}=\gamma_\mu$, $\mu=1,2,3$ for the vector meson.  
The baryon correlation functions are measured analogously.
%For baryons it is
%\begin{equation}
%  G_{\mathcal{O}}(t) = \sum_{x,y}\langle \bar{B}(x,t)B(y,0) \rangle,
%\end{equation}
%where 
%\begin{equation}
%  B(x,t)=\epsilon_{\alpha\beta\gamma}[D^{\alpha \rm{T}}(x,t)\gamma_0\gamma_2\gamma_5U^\beta(x,t)]U^\gamma(x,t)
%\end{equation}
%for spin 1/2 baryon (proton) and 
%\begin{align}
%  B(x,t)=&2\epsilon_{\alpha\beta\gamma}[D^{\alpha \rm{T}}(x,t)\gamma_0\gamma_2i(\gamma_1-i\gamma_2U^\beta(x,t)]U^\gamma(x,t)\\
%         &\epsilon_{\alpha\beta\gamma}[U^{\alpha \rm{T}}(x,t)\gamma_0\gamma_2i(\gamma_1-i\gamma_2U^\beta(x,t)]D^\gamma(x,t)
%\end{align}
%for spin 3/2 baryon ($\Delta_{3/2}$).

\begin{figure}
  \begin{center}
    \includegraphics[width=0.4\textwidth]{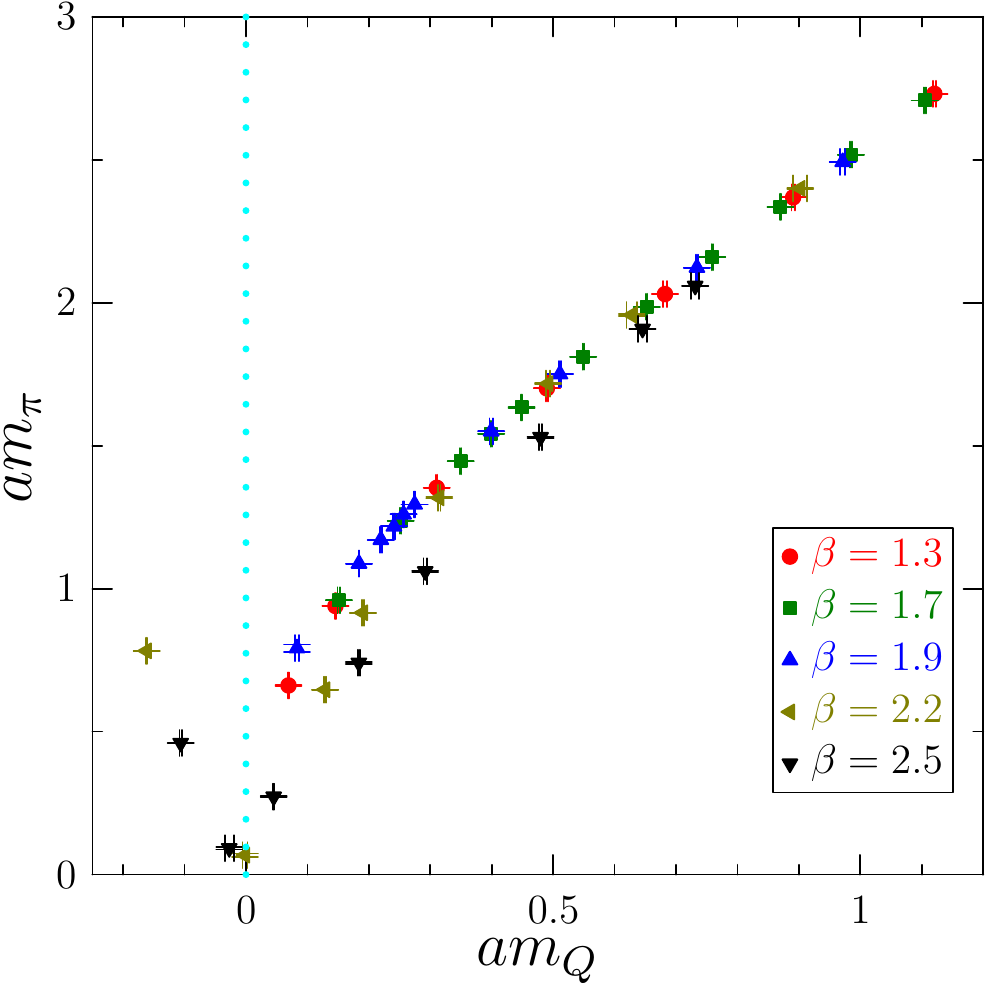}
    \includegraphics[width=0.4\textwidth]{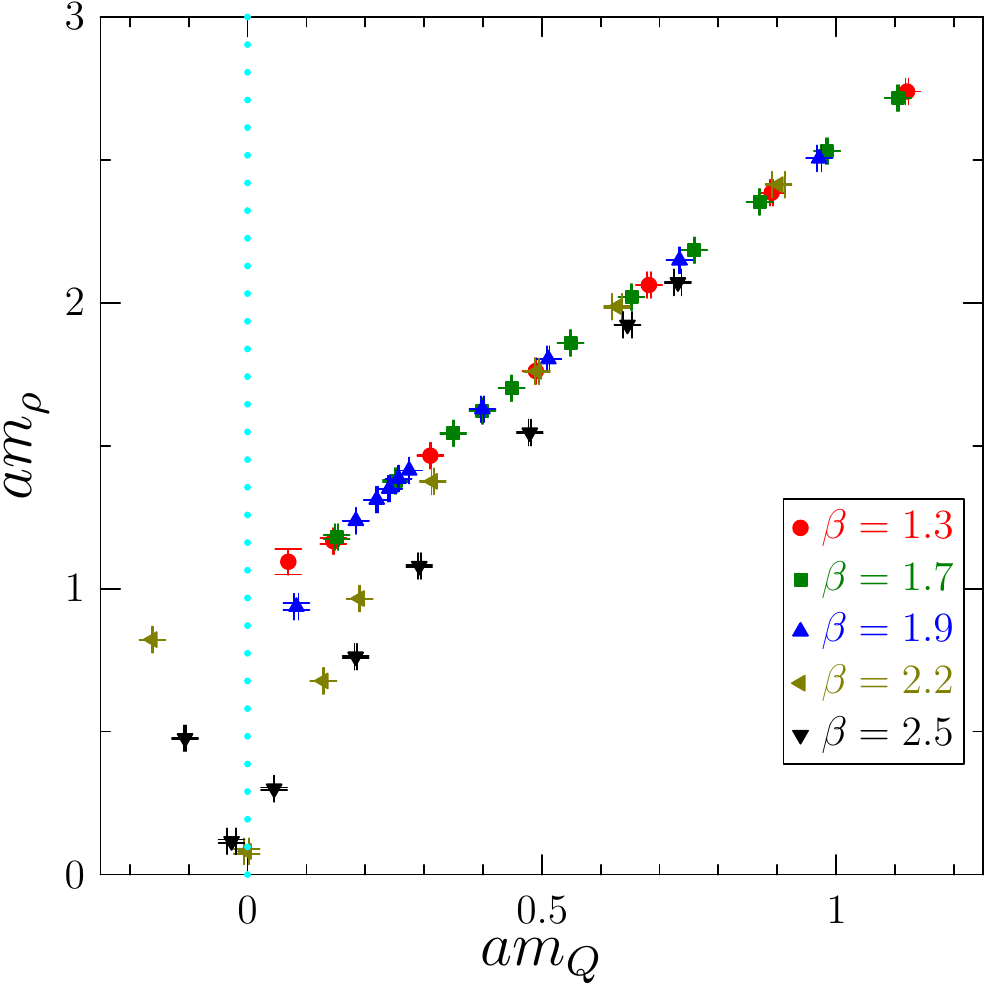}
    \caption{Mass of $\pi$, (pseudoscalar) and mass of $\rho$ (vector meson) with different $\beta$ as a function of PCAC quark mass. \label{mesons}}
  \end{center}
\end{figure}  

\begin{figure}
  \begin{center}
    \includegraphics[width=0.4\textwidth]{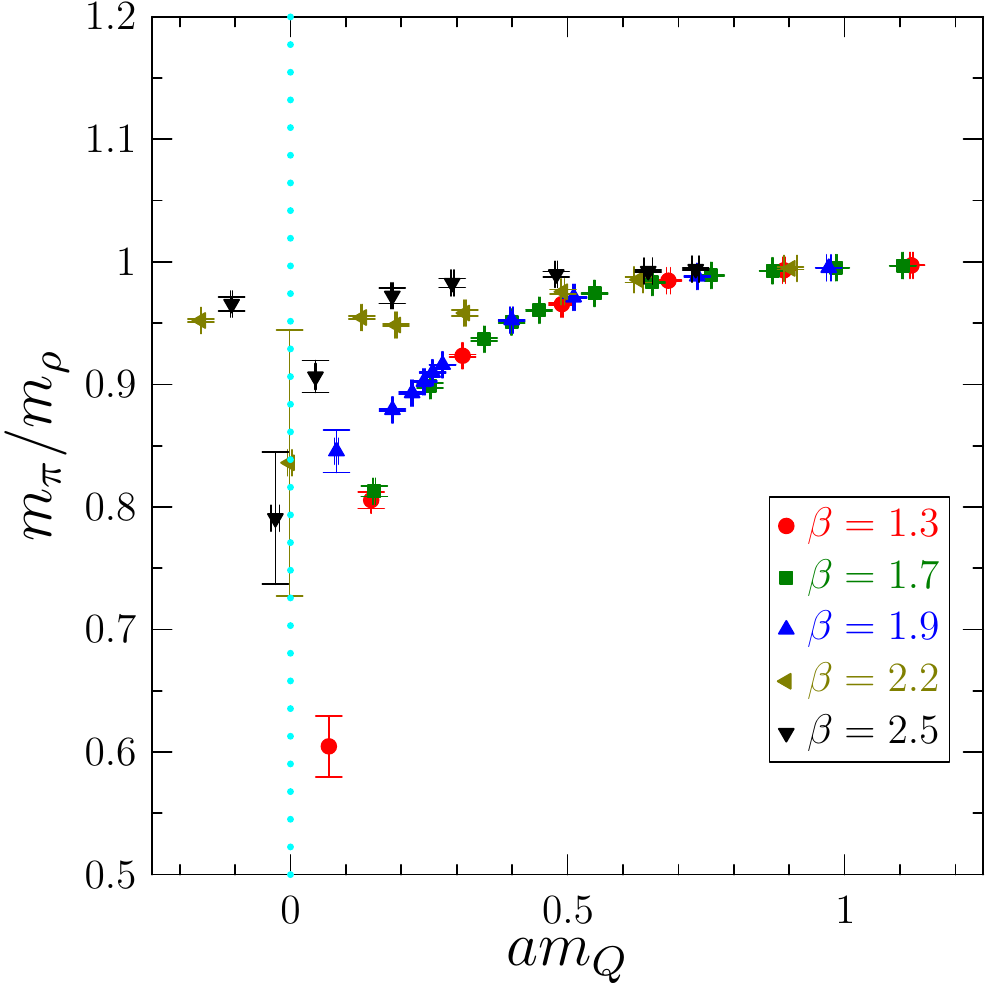}
    \includegraphics[width=0.4\textwidth]{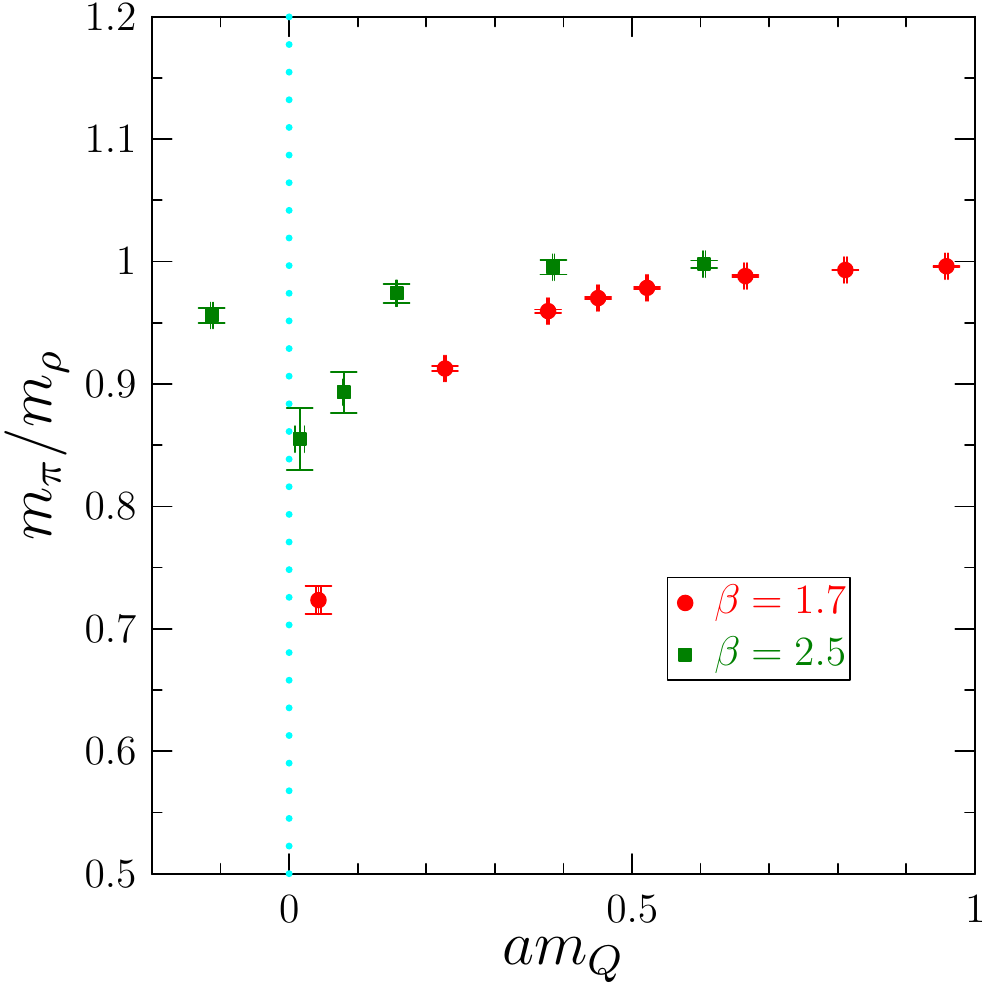}
    \caption{Ratio between $\pi$ (pseudoscalar) and $\rho$ (vector meson) mass with different $\beta$ as a function of PCAC quark mass. In the left panel the results are for adjoint representation and in the right panel for the fundamental representation.\label{mesonratios}}
  \end{center}
\end{figure}  

\begin{figure}
  \begin{center}
    \includegraphics[width=0.45\textwidth]{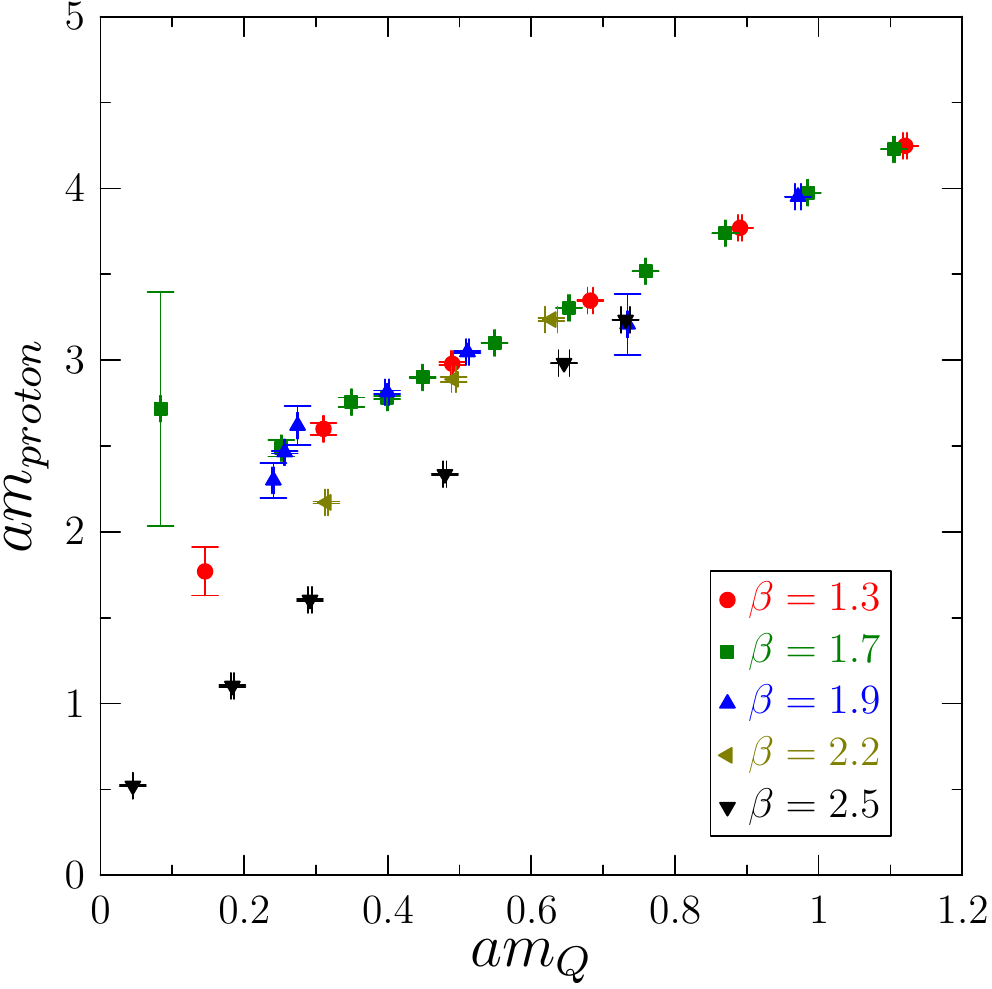}
    \includegraphics[width=0.45\textwidth]{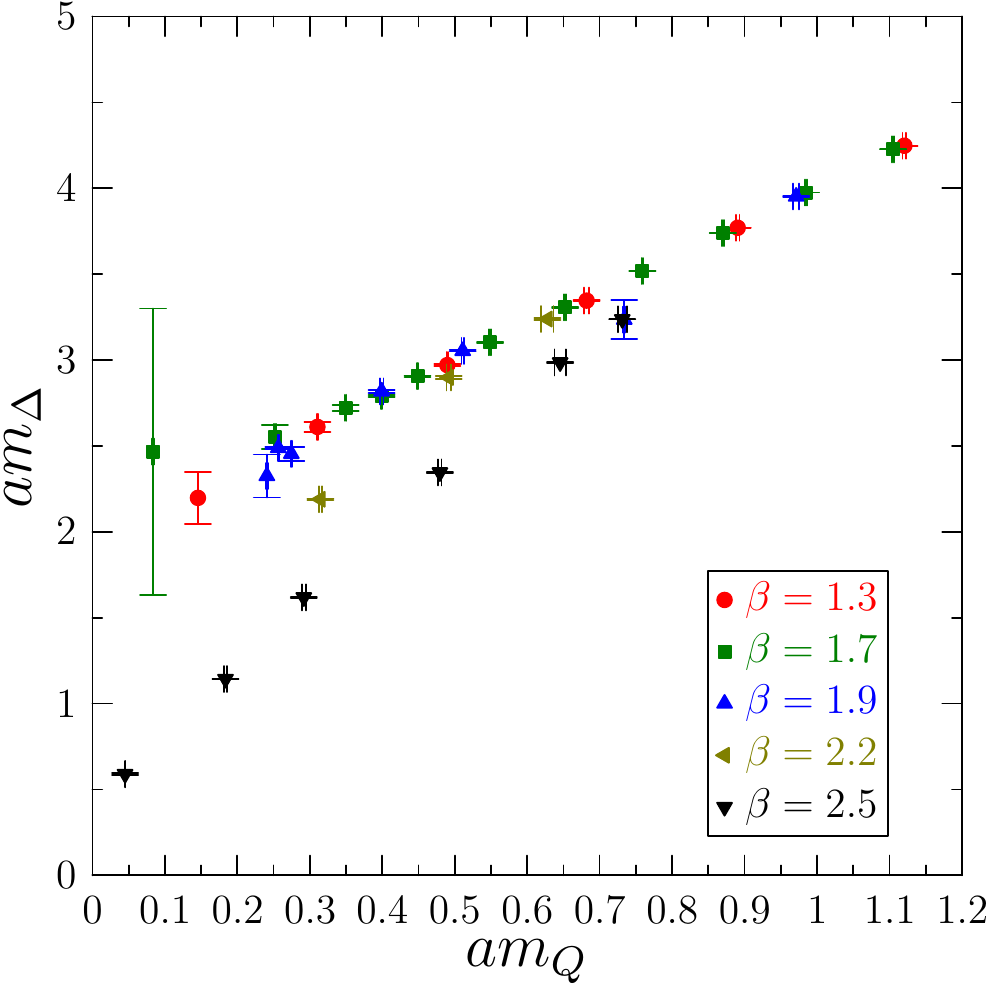}
    \caption{Mass of proton (spin 1/2 baryon) and $\Delta$ (spin 3/2 baryon) as a function of PCAC quark mass. \label{baryons}}
  \end{center}
\end{figure}

The masses of the pseudoscalar and vector mesons are plotted 
in Figs.~\ref{mesons} and \ref{mesonratios}. For a conformal 
theory one expects that all particle masses approach zero as 
$m_{\rm{Q}}\rightarrow0$, with the same exponent.  
However, at small $\beta$ we observe a more or less standard pattern
of chiral symmetry breaking: as $m_{\rm Q} \rightarrow 0$ the
pseudoscalar meson (Goldstone bosons for chiral symmetry breaking)
mass behaves approximately as $\propto\sqrt{m_{\rm Q}}$ at small
$m_{\rm Q}$, whereas vector meson remains massive.  On the other hand,
at large $\beta \gsim 2$ the mesons are practically degenerate, and
their masses almost vanish as $m_{\rm Q}\rightarrow 0$.  The baryon
masses are shown in Fig.~\ref{baryons} and show a similar pattern: at
small $\beta$ they extrapolate to a finite value as $m_{\rm
  Q}\rightarrow 0$, but at large $\beta$ the masses decrease linearly,
but with a small intercept.

Thus, one can argue that at large $\beta$ the results are compatible
with a conformal behaviour (if we ignore the small intercepts in
particle masses as $m_{\rm Q}\rightarrow 0$).  However, we emphasize
that this is qualitatively also compatible with standard QCD-like
running coupling: we have also measured the meson spectrum in theory
with SU(2) gauge + 2 fundamental representation fermions, and the mass
pattern is comparable to the one shown in Fig.~\ref{mesons}.  On
Fig.~\ref{mesonratios} we show the mass ratios of pseudoscalar and
vector mesons for adjoint and fundamental fermions.  For fundamental
fermions the reason for this behaviour is easy to understand: at small
$\beta$ we observe chiral symmetry breaking, as we should, but at
large $\beta$ the lattice spacing becomes so small that the linear
size of the system becomes much smaller than the hadron size.  Thus,
the quarks become effectively deconfined, and we observe
near-conformal behaviour also with fundamental quarks at $m_{\rm Q}
\approx 0$.

Thus, whether or not the adjoint quark theory has conformal or
near-conformal behaviour (and hence an IR fixed point or ``walking''
coupling), or QCD-like running coupling, is very difficult to
distinguish from the mass spectrum.  However, we note that if there is
a genuine IR fixed point where the theory becomes conformal, there
must be a phase transition somewhere along the critical $m_{\rm
  Q}=0$-line ($\kappa_c(\beta)$) where the theory goes from the
chirally broken phase (at small $\beta$) into the phase controlled by
the IR fixed point (at large $\beta$), where IR physics is conformal.
In our simulations we do observe behaviour compatible with this:
there is a clear change in the $m_{\rm Q}\rightarrow 0$
limit around $\beta \approx 2$.  Indeed, around this point it is very
difficult to even reach small $m_{\rm Q}$ values with lattice Monte
Carlo.  If this scenario is the correct one, then the chiral symmetry
breaking is a lattice artifact (or rather, UV cutoff artifact) 
not present in the continuum theory.  The existence of a critical 
point was also suggested in \cite{Catterall:2008qk}.

\section{Conclusions}
We have presented preliminary results of the lattice measurement 
of the mass spectrum in SU(2) gauge theory with two fermions in the adjoint 
representation.  This theory has been proposed to have
either a walking (i.e. very slowly evolving) coupling or even
an IR fixed point, where the theory becomes conformal (in the massless fermion
limit).  In this case the physical states of the theory are all massless.
We indeed observe an almost-massless behaviour at large inverse lattice
coupling $\beta$.  However, resolving the reason for this behaviour on 
the lattice is complicated by the fact that in practice the spectrum 
for a theory with a QCD-like running coupling also appears conformal 
at large $\beta$, due to the fact that the lattice volume becomes
so small that the system is essentially deconfined.  In order to resolve
the issue more direct evaluation of the evolution of the coupling
is required, using e.g. Schr\"odinger functional methods.

\section{Acknowledgements}
The simulations were performed on center of scientific computing Finland (CSC) and J\"ulich supercomputing center (JSC).
JR and KR acknowledge support from Academy of Finland grant number 114371.

\end{document}